\documentclass{aa}  

\usepackage{graphicx}
\usepackage{txfonts}
\usepackage{hyperref}
\usepackage{placeins}
\defcitealias{2023Kannan-MTNG}{K23}

\usepackage{ulem}
\begin{document} 

   \title{Investigating cosmic strings using large-volume hydrodynamical\\
   simulations in the context of \textit{JWST}'s massive UV-bright galaxies}
   \titlerunning{Investigating cosmic strings using large-volume hydrodynamical simulations}

    \author{Sonja M.~Koehler\inst{1, 2, 3}\thanks{Corresponding author; \href{mailto:sonja.koehler@cfa.harvard.edu}{\texttt{sonja.koehler@cfa.harvard.edu}}}
        \and
        Hao Jiao\inst{4, 5}
        \and
        Rahul Kannan\inst{6}\thanks{\email{\href{mailto:kannanr@yorku.ca}{\texttt{kannanr@yorku.ca}}}}
        }
    \authorrunning{S.~M.~Koehler et al.}

   \institute{Institute for Theoretical Physics, Goethe University, D-60438 Frankfurt am Main, Germany
        \and
            Department of Physics, Kavli Institute for Astrophysics and Space Research, Massachusetts Institute of Technology, Cambridge, MA 02139, USA
        \and
            Center for Astrophysics | Harvard \& Smithsonian, 60 Garden Street, Cambridge, MA 02138, USA
        \and
            Department of Physics, McGill University, Montr{\'e}al, QC, H3A 2T8, Canada
        \and
            Cosmology, Gravity and Astroparticle Physics Group, Center for Theoretical Physics of the Universe,
            Institute for Basic Science, Daejeon 34126, Korea
        \and
            Department of Physics and Astronomy, York University, 4700 Keele Street, Toronto, ON M3J 1P3, Canada
             }

   \date{Received X; accepted Y}

    \abstract{Recent observations from the \textit{James Webb Space Telescope} (\textit{JWST}) have uncovered an unexpectedly large abundance of massive, UV-bright galaxies at high redshifts, presenting a significant challenge to established galaxy formation models within the standard $\Lambda$CDM cosmological framework. Cosmic strings, predicted by a wide range of particle physics theories beyond the Standard Model, provide a promising potential explanation for these observations. They may act as additional gravitational seeds in the early universe, enhancing the process of high-redshift structure formation, potentially resulting in a more substantial population of massive, efficiently star-forming galaxies. We numerically investigate this prediction in large-volume hydrodynamical simulations using the moving-mesh code \textsc{arepo} and the well-tested IllustrisTNG galaxy formation model. We evaluate the simulation results in the context of recent \textit{JWST} data and find that sufficiently energetic cosmic strings produce UV luminosity and stellar mass functions that are in slightly to substantially better agreement with observations at high redshifts. Moreover, we observe that the halos seeded by cosmic strings exhibit a greater efficiency of star formation and enhanced central concentrations. Interestingly, our findings indicate that the simulations incorporating cosmic strings converge with those from a baseline $\Lambda$CDM model by redshift $z \sim 6$. This convergence suggests that the modified cosmological framework effectively replicates the successful predictions of the standard $\Lambda$CDM model at lower redshifts, where observational constraints are significantly stronger. Our results provide compelling evidence that cosmic strings may play a crucial role in explaining the galaxy properties observed by \textit{JWST} at high redshifts while maintaining consistency with well-established models at later epochs.}


   \maketitle

\section{Introduction}
\label{sec:Introduction}

The launch of the \textit{James Webb Space Telescope} \citep[\textit{JWST};][]{2006Gardner-JWST} has marked the beginning of a new era of precision observations at high redshifts. Recent surveys have revealed a surprisingly large number density of highly UV-luminous (e.g.~\citealt{2023aHarikane, 2024bHarikane, 2025Harikane, 2023Bouwens, 2024Adams, 2024Donnan, 2024Finkelstein, 2024McLeod, 2024Robertson}; though see also \citealt{2024Willott}), massive \citep[e.g.][]{2024Weibel, 2024Chworowsky} high-redshift galaxies, significantly exceeding almost all predictions from established galaxy formation models \citep[e.g.][]{2018Tacchella, 2019Yung, 2020Behroozi, 2020Vogelsberger-IllTNG_JWST, 2022Kannan-THESAN, 2023Kannan-MTNG}. Additionally, some observations have suggested the existence of potentially overmassive galaxies in the early universe \citep[e.g.][]{2023Labbe, 2023Akins, 2024Xiao, 2024Carniani, 2024Casey}.

The majority of these studies rely on photometrically selected galaxy samples, which are subject to uncertainties in spectral energy distribution (SED) fits \citep[e.g.][]{2023Steinhardt, 2023Endsley} and potentially sensitive to low-redshift interlopers \citep[e.g.][]{2022Naidu, 2023aFujimoto, 2023Zavala}. However, spectroscopic follow-ups  \citep[e.g.][]{2023Curtis-Lake, 2023aRobertson, 2023Wang, 2023bFujimoto, 2023Bunker, 2023Bakx, 2023aArrabalHaro, 2024Castellano, 2024DEugenio} have generally found good agreement with the photometrically inferred redshifts and UV luminosity functions \citep[UVLFs; e.g.][]{2024Finkelstein, 2024bHarikane, 2025Harikane, 2025Naidu}.

These observations' compatibility -- or lack thereof -- with the standard Lambda Cold Dark Matter ($\Lambda$CDM) model of cosmology has been discussed extensively \citep[e.g.][]{2023Boylan-Kolchin, 2023Biagetti, 2023Ferrara, 2023Finkelstein, 2023Lovell, 2023Mason, 2023Steinhardt, 2024Desprez, 2024Yung} since the first publication of \textit{JWST}'s early science observations. Many compelling potential explanations for these results have been proposed in the literature to date. Examples within standard $\Lambda$CDM cosmology include 
(i)~a top-heavy initial mass function (IMF) at high redshifts \citep[e.g.][]{2022Inayoshi, 2023Finkelstein, 2023aHarikane, 2023Steinhardt, 2024Yung, 2025Harvey, 2025Lu},
(ii)~a significantly increased star formation efficiency in high-redshift galaxies \citep[e.g.][]{2023Dekel, 2024Finkelstein, 2024Ceverino, 2024Chworowsky, 2025Boylan-Kolchin},
(iii)~field-to-field statistical variance \citep[e.g.][]{2024Desprez, 2024Willott}{}{},
(iv)~an increased UV variability due to a bursty star-formation history \citep[SFH; e.g.][]{2023Shen, 2023aSun, 2023bSun, 2023Pallottini, 2023Munoz, 2024Casey},
(v)~a lack of dust attenuation at high redshifts \citep[e.g.][]{2023Ferrara, 2023Mason, 2023Finkelstein}, and
(vi)~contributions from active galactic nuclei \citep[AGN; e.g.][]{2022Inayoshi, 2023bHarikane, 2024Hedge, 2024Chworowsky}.

Additionally, a number of proposed modifications to the $\Lambda$CDM paradigm have been argued to potentially alleviate this tension \citep[e.g.][]{2022Liu, 2022Menci-DDE, 2023Boylan-Kolchin, 2023JiaoHao-JWST, 2023Biagetti, 2023Gong-FDM, 2023Huetsi-axions_PBH, 2024Shen-EDE, 2025Shen_EDE, 2025KumarWeiner}. A notable example is the cosmic string model, with predictions of a large abundance of massive, efficiently star-forming galaxies at $z \gtrsim 10$ predating \textit{JWST}'s first light by several years \citep[e.g.][]{2012Shlaer}.\footnote{Earlier investigations of cosmic strings recognized their role as potential seeds for structure formation several decades prior, see e.g. \citet{1976Kibble, 1986Rees, 1986TurokBrandenberger}. However, this earlier work focuses on a highly energetic string model now ruled out by cosmic microwave background observations \citep{2014PlanckCol-CMB_constraints, 2016Charnock-CMB_constraints}.}
Cosmic strings are linear topological defects predicted by a broad range of beyond the Standard Model (BSM) particle physics theories. Topological defects form in the early universe if the vacuum manifold after a symmetry-breaking phase transition has a non-trivial homotopy group; if this manifold has the topology of a circle, the defects are one-dimensional strings carrying trapped energy.

The simplest model of stable cosmic strings, i.e. the Nambu-Goto model\footnote{There are other cosmic string models with more parameters, including string theory models \citep[so-called cosmic superstrings, e.g.][]{1985Witten, 2005Kibble}, the Abelian Higgs model \citep{1973Nielsen-HiggsCS}, and superconducting cosmic string models \citep{1985Witten}.} \citep[e.g.][]{2000VilenkinShellard-book}, is fully characterized by only one free parameter: the string tension $\mu$ describing the cosmic strings' energy per unit length. The string tension is related to the energy scale $\eta$ of the symmetry breaking phase transition as $\mu \sim \eta^2$ and typically expressed in terms of the dimensionless\footnote{We use natural units in this work, $c=k_B=\hbar=1$.} string tension $G\mu$, where $G$ is Newton's constant of gravitation.
For detailed reviews on cosmic strings, their proposed formation in specific field theories, as well as their spatial distribution and scaling, see e.g. \citet{1994Brandenberger-review, 1995HindmarshKibble, 2000VilenkinShellard-book, 2002Durrer}.

By causality, a network of long cosmic strings and loops will inevitably form in the expanding universe and persist until today in particle physics models admitting cosmic string solutions \citep{1980Kibble, 1982Kibble}. Cosmic string loops are generated by the intersections of long strings. A “scaling solution” of the network is approached rapidly after the phase transition, according to which all statistical properties of the network are time-invariant if all distances are scaled to the Hubble radius \citep{1992CS-Scaling}. This is verified by Nambu-Goto simulations \citep{1993CS-Scaling, CSsimuls1, CSsimuls2, CSsimuls3, CSsimuls4, CSsimuls5, CSsimuls6, CSsimuls7, CSsimuls8, CSsimuls9, CSsimuls10}. However, while the scaling solution of long strings is robust, as it is determined by causality arguments, the scaling distribution of string loops is less certain, being affected by the decay channels of both long strings and loops.

Therefore, the most robust constraints on the string tension come from the signals of long strings, the strongest one being the upper limit $G\mu<10^{-7}$ inferred from cosmic microwave background (CMB) anisotropies \citep{2016Charnock-CMB_constraints, CS-CMB-2, CS-CMB-3}. With the assumption that cosmic string loops only lose energy to gravitational radiation, a tighter upper bound $G\mu \lesssim 10^{-10}$ can be inferred from the stochastic gravitational wave background recently detected by millisecond pulsar timing arrays \citep[PTAs;][]{CS-NANO-1,NANOGrav2023,CS-NANO-Wang}, but we emphasize that this constraint is less robust due to the uncertainty in the scaling solution of loops.
Additionally, we note that the gravitational wave signal is dominated by the smallest loops, while larger loops contribute most significantly to galaxy formation \citep[e.g.][]{2024JiaoHao-N-body}. Thus, we adopt only the robust constraint on the string tension in this work.

Since cosmic string loops are predicted to act as additional gravitational seeds in the early universe -- alongside the density fluctuations described by standard $\Lambda$CDM cosmology -- they provide a well-motivated explanation for enhanced early structure formation and therefore a larger abundance of massive dark matter halos at high redshifts \citep[e.g.][]{1994Brandenberger-review, 1996Moessner, 2000VilenkinShellard-book}. This, in turn, is expected to increase the abundance of massive high-redshift galaxies, such as the ones observed by \textit{JWST} \citep{2023JiaoHao-JWST}.\footnote{We note that a further prediction of these models is the existence of early supermassive black holes \citep[SMBHs;][]{2015Bramberger, 2022Cyr}, another highly-debated source of friction between standard galaxy formation models and high-redshift observations by \textit{JWST} and prior surveys \citep[see e.g.][]{2010Volonteri, 2015Wu, 2022Pacucci, 2024Silk, 2024Greene}, as well as intermediate-mass black holes \citep[IMBHs;][]{2021Brandenberger-IMBH}, for which strong observational evidence has been found in recent years \citep{2020Abbott-LIGO-IMBH, 2024Haeberle-IMBH}.}

To investigate this theoretical prediction, as well as the impact of modeling cosmic strings on the high-redshift UVLF, we present results from large-volume ($L_{\text{box}}=148 \,{\rm cMpc}$) hydrodynamical simulations using the finite-volume moving-mesh code \textsc{arepo} \citep{2010Springel-AREPO, 2016Pakmor-AREPO} and the well-tested IllustrisTNG \citep{2018Springel-IllTNG, 2018Nelson-IllTNG, 2019Nelson-IllTNG, 2018Pillepich-IllTNG, 2018Marinacci-IllTNG, 2018Naiman-IllTNG} galaxy formation model. This is further building on previous work presented in \citet{2024JiaoHao-N-body} of lower-resolution $N$-body simulations focusing on the high-redshift halo mass functions (HMFs).
A crucial advantage of full-physics hydrodynamical simulations in this context is the ability to self-consistently model not only the HMFs, but also the stellar mass functions (SMFs) and UVLFs. Rather than inferring the latter from HMFs of dark matter-only simulations, this directly takes baryonic feedback processes into account, allowing more accurate predictions for direct comparisons to observational data.

This work is organized as follows. We describe our methodology in Sect.~\ref{sec:Methods}. The main results, including comparisons to \textit{JWST} observations and existing $\Lambda$CDM simulation predictions, are presented in Sect.~\ref{sec:Results}. Finally, we discuss and summarize our findings in Sect.~\ref{sec:Discussion_and_Conclusion}.

\section{Methods}
\label{sec:Methods}

An overview of the simulations presented in this work is given in Table~\ref{tab:simulations_overview}. In summary, we run a baseline simulation in standard $\Lambda$CDM cosmology, three runs modeling the impact of cosmic string loops with string tension $G\mu = 10^{-8}$ (CS-8), and three $G\mu = 10^{-10}$ runs (CS-10) using the same relative distribution of cosmic string loops as the CS-8 runs. All simulations are based on the same $\Lambda$CDM initial conditions (ICs), but to improve the statistics of our results, we employ multiple CS-8 and CS-10 runs using different IC modifications described below.

\begin{table}[t]
	\centering
    \caption{Overview of the simulations.}
	\label{tab:simulations_overview}
    \begin{tabular}{l l l}
        \hline \hline
		Name & $G\mu$ & Number of runs \\
		\hline
		$\Lambda$CDM & -- & 1 \\
		CS-8 & $10^{-8}$ & 3 (CS-8-0, CS-8-1, CS-8-2) \\
        CS-10 & $10^{-10}$ & 3 (CS-10-0, CS-10-1, CS-10-2) \\
        \hline
	\end{tabular}
    \tablefoot{We run one baseline simulation using concordance $\Lambda$CDM cosmology and three runs modeling cosmic string loops for each of the string tension values $G\mu=10^{-8}$ and $G\mu=10^{-10}$. Runs labeled CS-8-$x$ and CS-10-$x$ with the same $x$ share the same spatial and radius distribution of cosmic string loops, as well as the same relative mass distribution scaled to the respective value of the string tension.}
\end{table}

The $\Lambda$CDM ICs were generated at redshift $z=127$ using the simulation code \textsc{gadget-4} \citep{2021Springel-GADGET4}, a recent update to the $\textsc{gadget}$ code base \citep{2001Springel-GADGET}. They model the initial dark matter distribution using second-order Lagrangian perturbation theory in a periodic, cubic box with side length $L_{\rm{box}} = 100 \,h^{-1}\,{\rm cMpc} = 148 \,{\rm cMpc}$, resolved by $850^3$ dark matter particles.
The cosmological parameters employed throughout all simulations presented in this work are the following: $\Omega_{\rm m} = 0.3089$, $\Omega_{\rm b} = 0.0486$, $\Omega_{\Lambda} = 0.6911$, $H_0 = 100\,h\,\,{\rm km}\,{\rm s}^{-1} \,{\rm Mpc}^{-1} = 67.74\,\,{\rm km}\,{\rm s}^{-1} \,{\rm Mpc}^{-1}$, $\sigma_{8} = 0.8159$, and $n_{\rm s} = 0.9667$ \citep{2016Planck}.

For the runs including cosmic strings, these ICs are then modified at the initial redshift by computing the changes in dark matter particle positions and velocities due to the gravitational effect of cosmic string loops using the Zel’dovich approximation (\citealt{1970Zeldovich}; see also Appendix of \citealt{2024JiaoHao-N-body}).
This modification is done separately for each of the runs; the specifics of this step are described in more detail in Sect.~\ref{subsec:Implementation_cosmic_strings}.

For each of the modified ICs, as well as the unmodified ICs generated directly with \textsc{gadget-4}, we then separately run full-physics hydrodynamical simulations down to redshift $z=6$ using the finite-volume moving-mesh code \textsc{arepo}. The $2 \times 850^3$ dark matter and gas particles in the box volume of $(148 \,{\rm cMpc})^3$ correspond to dark matter and gas mass resolutions of $m_{\rm DM} = 1.74 \times 10^8 {\,\rm M_\odot}$ and $m_{\rm gas} = 3.24 \times 10^7 {\,\rm M_\odot}$, respectively. The gravitational softening length for both dark matter and star particles is set to $4.4 \,{\rm ckpc}$. For gas in the simulation volume, the softening is locally adaptive \citep{2010Springel-AREPO}, i.e. variable depending on cell size, with a minimum softening length of $0.44 \,{\rm ckpc}$.

We use the IllustrisTNG \citep[e.g.][]{2018Springel-IllTNG, 2019Nelson-IllTNG} galaxy formation model, which is an updated version of the Illustris \citep{2013Vogelsberger-Ill, 2014bVogelsberger-Ill, 2014aVogelsberger-Ill, 2014Torrey-Ill} model and has been shown to accurately reproduce a wide range of properties of the observed low-redshift galaxy population \citep[e.g.][]{2018Nelson-IllTNG, 2021Nelson-IllTNG, 2018Springel-IllTNG, 2018Marinacci-IllTNG, 2018Naiman-IllTNG}. Particularly relevant for this work is the model's ability to produce SMFs consistent with observations \citep{2018Pillepich-IllTNG}, as well as realistic star-formation activity \citep{2019Donnari-IllTNG} at low redshifts. Further, the coupling of this physics model to the radiation hydrodynamics (RHD) extension \textsc{arepo-rt} \citep{2019Kannan-AREPO-RT} in the context of the reionization simulation \textsc{thesan} \citep{2022Kannan-THESAN, 2022bKannan, 2022Garaldi, 2022Smith, 2024Garaldi-THESAN} has been shown to make accurate predictions for numerous galaxy population properties at redshifts $5 \lesssim z \lesssim 10$, including the UVLFs \citep{2022Kannan-THESAN}.
However, at very high redshifts -- in particular beyond $z \gtrsim 12$ -- the model, based on concordance $\Lambda$CDM cosmology, underpredicts the number density of highly UV-luminous galaxies and their star formation rates (\citealt{2023Kannan-MTNG}, in the following \citetalias{2023Kannan-MTNG}). In this work, the results from the combined data of several large simulation efforts \citepalias{2023Kannan-MTNG} based on this model are used as an additional baseline $\Lambda$CDM comparison, as well as to determine resolution corrections for the simulations presented in this work (see Sect.~\ref{subsec:Resolution_correction}).

Halo and galaxy masses in each of our simulations are determined using the \textsc{subfind} substructure finder \citep{2001Springel-SUBFIND} based on the friends-of-friends \citep[FOF;][]{1985Davis-FOF} group finder within \textsc{arepo}. Specifically, the newer variant \textsc{subfind-hbt} based on Hierarchical Bound-Tracing \citep[HBT;][]{2012Han-HBT, 2018Han-HBT} is employed on-the-fly to extract information about substructure in the simulation volume and its properties at each of the output redshifts $z=6, 7, 8, \dots, 19$.

\subsection{Implementation of cosmic string loops in the initial conditions}
\label{subsec:Implementation_cosmic_strings}

We modify the simulation's ICs to include cosmic strings based on the procedures described in \citet{2024JiaoHao-N-body}. The detailed steps used in this work are as follows.

To obtain the masses and positions of cosmic string loops, we sample the loop radii based on the scaling distribution and randomly assign their positions separately in the three runs with string tension $G\mu=10^{-8}$, labeled as CS-8-0, CS-8-1, and CS-8-2 (Table~\ref{tab:simulations_overview}). The number density of loops per unit radius is given in Appendix~\ref{app:Distribution_CS}. The mass distribution follows from the radius distribution, as the mass of a loop with radius $R$ is given by $M_{\rm loop}=\beta\mu R$, where $\beta R$ describes the mean length of a loop with radius $R$ (see e.g. \citealt{1994Brandenberger-review} for details).
We adopt a lower cutoff of $10^7\, {\rm M_\odot}$ ($10^5\, {\rm M_\odot}$) for the loop mass in the CS-8 (CS-10) runs due to the finite computational resources, ensuring that this cutoff is set lower than both the dark matter and gas mass resolution in the simulations.

To allow for a direct comparison, we use these same relative loop distributions for the three CS-10 runs by scaling down the CS-8 masses by a factor of $1/100$, while keeping their positions and radii identical. The mass scaling factor is based on the fact that the loop mass is proportional to the string tension. These runs are analogously labeled CS-10-0, CS-10-1, and CS-10-2, where the same final digit $x$ indicates the same relative distributions of loops in the respective CS-8-$x$ and CS-10-$x$ runs.
Using the different distributions, we verify that this sampling procedure based on random number generation in the numerical implementation does not cause substantial variations in the mass functions of our simulated halo population (Appendix \ref{app:Distribution_CS}).

We modify the positions and velocities of dark matter particles according to the gravitational displacement onto cosmic string loops, calculated using the Zel'dovich approximation \citep[for further details, see also][]{2024JiaoHao-N-body, 2024JiaoHao-Accretion}.
For this step and the subsequent full-physics runs, the following approximations were further made. First, cosmic string loops are treated as point masses, i.e. their radius is considered to be negligible on the scales of interest. This is based on the fact that except for very few ($\mathcal{O}(1)$ in each simulation) of the largest loops, the radii of the vast majority of the $9\,158$ cosmic string loops in the simulation box are smaller than the mean separation of dark matter particles.
Second, the impact of cosmic string loops is only modeled by the IC modification, i.e. the masses of the loops themselves are neglected during the subsequent evolution of the simulation. We base this approximation on the fact that the total mass of cosmic string loops corresponds to only a very small fraction ($\mathcal{O}(10^{-6})$ for $G\mu=10^{-8}$) of the total mass of dark matter; further, the mass of a cosmic string loop does not significantly affect the evolution of a loop-seeded halo after the halo mass exceeds the loop mass \citep{1994Brandenberger-review}.

\subsection{Resolution corrections and baseline $\Lambda$CDM data}
\label{subsec:Resolution_correction}

We apply numerical resolution corrections to the stellar masses and UV magnitudes to mitigate the slightly lower star-formation efficiency caused by the relatively low effective resolution \citep[cf. e.g.][]{2018Pillepich-IllTNG} of our simulations. For this, we use the data presented in \citetalias{2023Kannan-MTNG}, based on the combination of results from the MillenniumTNG (MTNG; \citealt{2023Pakmor-MTNG, 2023Hernandez-Aguayo-MTNG}; \citetalias{2023Kannan-MTNG}) simulation MTNG740, the \textsc{thesan} flagship run $\textsc{thesan-1}$ \citep{2022Kannan-THESAN}, as well as the IllustrisTNG runs TNG50 \citep{2019Pillepich-TNG50, 2019Nelson-TNG50} and TNG300 \citep{2018Springel-IllTNG}.
All of these simulations and ours share the same code base, galaxy formation model, and \citet[]{2016Planck} cosmological parameters, while including both higher-resolution (TNG50, TNG300, \textsc{thesan-1}) and larger-volume (TNG300, MTNG740) runs, allowing more precise predictions over a larger mass and redshift range than our $\Lambda$CDM run alone.
For the same reason, the combined data additionally serve as a particularly useful $\Lambda$CDM baseline for comparison to our results (Sect.~\ref{sec:Results}).

Specifically, we implement the resolution corrections as outlined in \citetalias{2023Kannan-MTNG}, based on the methodology described in \citet{2018Pillepich-IllTNG}. This uses the highest-resolution run mentioned above, TNG50, as a baseline, while corrections for successively lower-resolution runs (in order, \textsc{thesan}, TNG300, MTNG740) are determined by computing the average differences in stellar mass between runs in overlapping halo mass bins (i.e., mass bins well-resolved by both simulations in each of these steps; see \citetalias{2023Kannan-MTNG} for further details). The resolution corrections for UV luminosities are computed analogously to these stellar mass corrections. We add the MTNG740 corrections as constant offsets to the logarithm of the stellar masses and to the UV magnitudes of all galaxies in our simulation. This is based on the fact that our runs have a quite similar effective resolution as the MTNG740 simulation, which has dark matter and gas mass resolutions of $m_{\rm DM} = 1.62 \times 10^8 {\,\rm M_\odot}$ and $m_{\rm gas} = 3.10 \times 10^7 {\,\rm M_\odot}$, respectively. To ensure self-consistency, allowing us to directly compare results from the simulations with and without cosmic string loops, the same offsets are applied to the $\Lambda$CDM run and all CS-8 and CS-10 runs, respectively.
Our output redshifts differ slightly from those considered in \citetalias{2023Kannan-MTNG}, i.e. $z_{\rm K23}=8$, 9, 10, 11, 12, 15. However, since the resolution corrections are only weakly dependent on redshift\footnote{The redshift-averaged correction value and its standard deviation is $(0.40 \pm 0.05) \, {\rm dex}$ for the logarithmic stellar masses $\log (M / {\rm M_\odot})$, and $(-1.32 \pm 0.14) \, {\rm mag}$ for the UV magnitudes.}, we use their average values as corrections for our additional output redshifts.
Finally, we further investigate the impact of numerical resolution in App.~\ref{app:resolution_tests}.

\section{Results}
\label{sec:Results}

In this section, we present our results for the halo mass functions (HMFs), stellar mass functions (SMFs), UV luminosity functions (UVLFs), stellar-to-halo mass relation (SHMR), and concentration-mass relation (cMr) from simulations with and without cosmic strings.
For the runs with cosmic string loops, we show the combined mass and luminosity functions $\phi_{\rm{combined}}$ inferred from all three runs, which are computed as weighted averages given by \citep{2020Vogelsberger-IllTNG_JWST}
\begin{equation}
    \phi_{\rm{combined}} = \frac{\sum_i \phi_i N_i^2}{\sum_i N_i^2} \thickspace,
    \label{eq:phi_combined}
\end{equation}
where $\phi_i$ refers to the mass/luminosity function of simulation $i \in \{ \text{CS-8-0, CS-8-1, CS-8-2} \}$ for $G\mu=10^{-8}$ or $i \in \{ \text{CS-10-0, CS-10-1, CS-10-2} \}$ for $G\mu=10^{-10}$ (cf. Table~\ref{tab:simulations_overview}). Here, $N_i$ is the number of halos (for the HMFs) or galaxies (for the SMFs and UVLFs) in the corresponding mass bin.

\subsection{Halo and stellar mass functions}
\label{subsec:Results_HMF_SMF}

\begin{figure*}
    \centering
    \includegraphics[width=17cm]{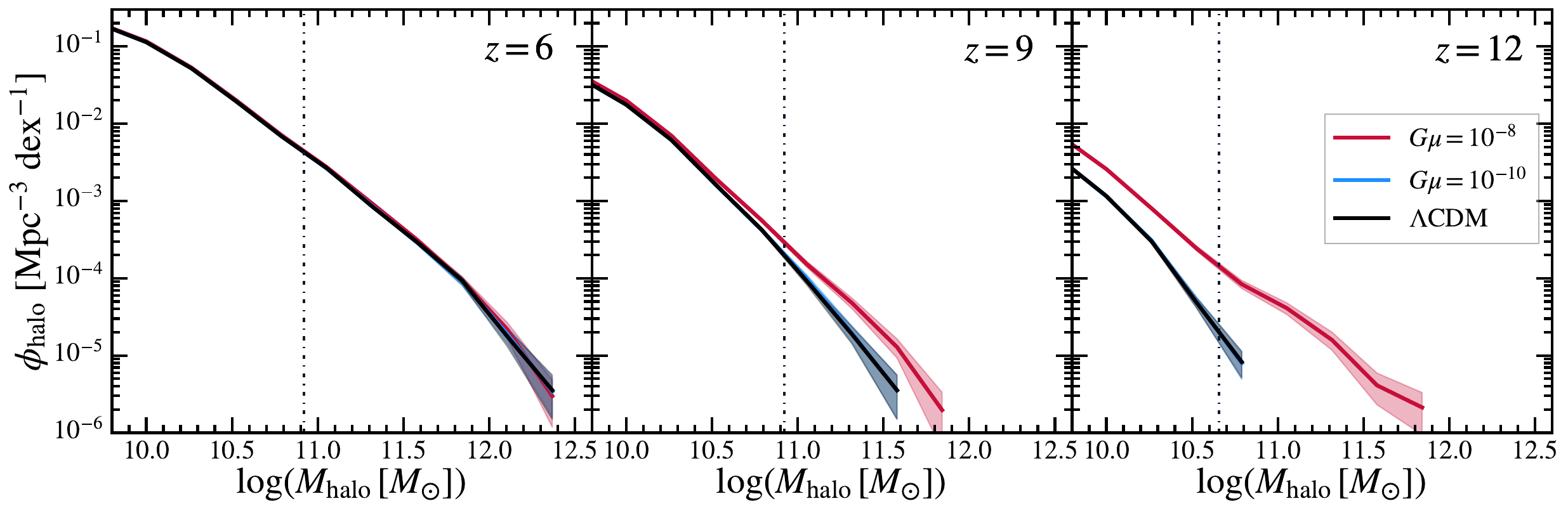}
    \caption{Comparison of halo mass functions from the $\Lambda$CDM run (solid black curves) to the runs modeling cosmic strings with string tension $G\mu=10^{-8}$ (solid red curves) and $G\mu=10^{-10}$ (solid blue curves) with corresponding Poisson errors (shaded error bands) at redshifts $z=6$, 9, and 12. Vertical dash-dotted lines indicate the first halo mass bin with 95\% completeness (see text for details). We note that these are color-coded analogously to the solid curves, but overlapping in the plots shown here.}
    \label{fig:HMF}
\end{figure*}

Figure~\ref{fig:HMF} shows the HMFs computed from our baseline $\Lambda$CDM run, as well as the runs in which the initial conditions have been modified by the effect of cosmic string loops with string tension $G\mu=10^{-8}$ and $G\mu=10^{-10}$ at redshifts $z=6$, 9, and 12. We additionally plot the Poisson errors of the mass functions and the lowest 95\% complete halo mass. The latter refers to the first halo mass bin in which at least 95\% of simulated halos have at least one stellar particle; halos with significantly lower masses than this are expected to be poorly resolved by the simulation. We use 20 logarithmically spaced halo mass bins spanning the mass range of $7.5 < \log (M_{\rm halo} / {\rm M_\odot}) < 12.5$.

By redshift $z=6$, our results from both the CS-8 and the CS-10 runs are converged with the $\Lambda$CDM baseline, showing that the impact of the modified cosmology on our HMFs becomes negligible towards lower redshifts.
However, as expected from the predictions of analytical mass functions \citep{2023JiaoHao-JWST}, the impact of cosmic string loops with $G\mu=10^{-8}$ increases towards higher redshifts.
For $z \gtrsim 9$, their inclusion results in a significantly larger abundance of halos at the high-mass end, as well as an overall slightly increased amplitude of the HMF.
At very early times, as shown in the $z=12$ panel, halos in the CS-8 runs both form more abundantly and grow substantially more massive than in the other runs. This results in an increased HMF amplitude across all mass ranges, with a particularly pronounced difference at the high-mass end.

\begin{figure*}
    \centering
    \includegraphics[width=17cm]{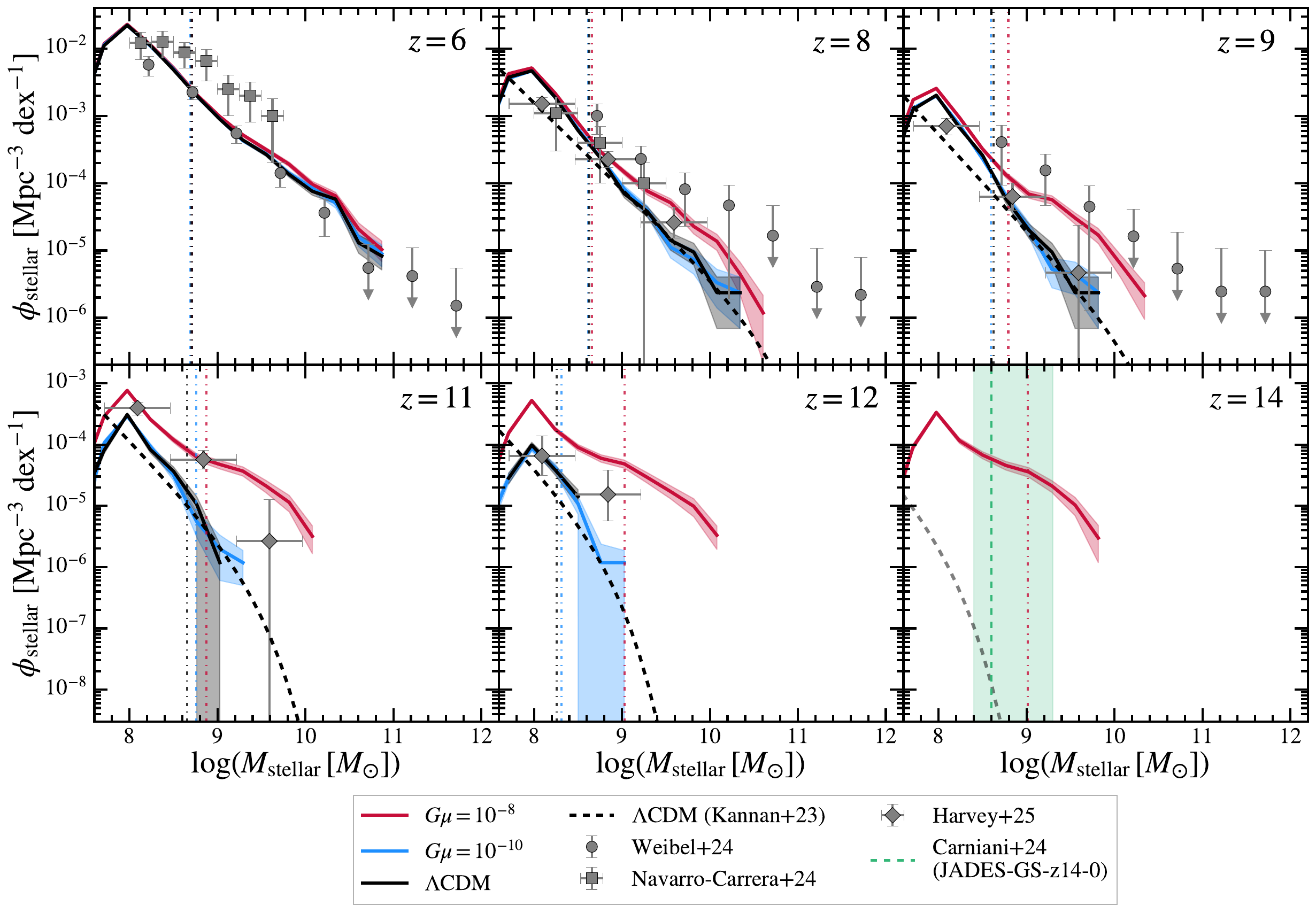}
    \caption{Simulated stellar mass functions from the $\Lambda$CDM run (solid black curves) and the runs modeling cosmic strings with string tension $G\mu=10^{-8}$ (solid red curves) and $G\mu=10^{-10}$ (solid blue curves) with Poisson errors (shaded error bands) at redshifts $z=6$ to $z=14$. Vertical dash-dotted lines in the corresponding colors show the median stellar mass of halos within the first 95\% complete halo mass bin to provide an indication of the typical minimum stellar mass of numerically well-resolved galaxies (see text). Dashed black curves show the Schechter fit of the simulated stellar mass functions from \protect\citetalias{2023Kannan-MTNG}, serving as an additional reference for standard $\Lambda$CDM cosmology from higher-resolution and larger-volume simulations using the same galaxy formation model as our runs. We show results from their closest available output redshift $z_{\rm K23} \sim 15$ in the $z=14$ panel, indicating the slight redshift discrepancy with a reduced opacity of the dashed curve. Observational \textit{JWST} data from \protect\citet{2024Weibel}, \protect\citet{2024Navarro-Carrera}, and \protect\citet{2025Harvey} are shown as gray symbols (at the closest integer redshift, where applicable). Additionally, the dashed green line in the lower right panel indicates the stellar mass of the spectroscopically confirmed massive $z \sim 14$ galaxy JADES-GS-z14-0 \protect\citep{2024Carniani}. The predicted number density from our CS-8 runs for such a galaxy at $z=14$ exceeds the $\Lambda$CDM prediction of \protect\citetalias{2023Kannan-MTNG} at their output redshift $z_{\rm K23} \sim 15$ ($z_{\rm K23} \sim 12$) by roughly 3.5 (1.5) orders of magnitude.}
    \label{fig:SMF}
\end{figure*}

However, we find that modeling cosmic strings with string tension $G\mu=10^{-10}$ has no significant impact on the HMF based on the resolved population of halos in our runs, instead showing excellent agreement with the $\Lambda$CDM run at all output redshifts.
We stress, though, that our simulations cannot fully capture all potential effects of $G\mu=10^{-10}$ loops: in particular, the most massive loop in the CS-10 runs has a mass of the order $m_{\rm CS, max} \sim 10^8 \, M_\odot$ (see App.~\ref{app:Distribution_CS}), but our simulations do not resolve the internal structure or even the abundance of halos on similar mass scales in a converged way, given our limited dark matter mass resolution. 
However, since our approach models the gravitational effects of cosmic strings through a modification of the initial conditions -- rather than as individual resolved mass particles (Sect.~\ref{subsec:Implementation_cosmic_strings}) -- we can still model these as additional gravitational seeds present at very early times ($z=127$). We therefore expect a potential impact on halos that eventually grow well beyond the maximum loop mass via the additional gravitational perturbations at early times. This is supported by the substantial impact of the $G\mu =10^{-8}$ loops on the HMF at halo masses $M_{\rm halo} > 10^{11.5}$ (Fig.~\ref{fig:HMF}), despite the maximum loop mass of $m_{\rm CS, max} \sim 10^{10} \, M_\odot$ in these runs, as well as by the minor differences we do observe between the $\Lambda$CDM and $G\mu=10^{-10}$ runs in Figs.~\ref{fig:SMF} and \ref{fig:UVLF} (discussed further in the following).
Nevertheless, there is a significant resolution limitation associated with these runs, as smaller loop-seeded halos themselves are not individually resolved. Based on our available data, we can only place a constraint on the \textit{resolved} halo and galaxy population (the scales corresponding to \textit{JWST} high-redshift observations), for which we only see a minimal impact by $G\mu=10^{-10}$ loops, while the impact of these loops is likely more substantial for halo masses below our resolution limit.

We also note that the structure finder did not identify any bound halos in the simulation volume for the $\Lambda$CDM and CS-10 runs at redshifts $z \geq 14$ at our resolution, while finding (sub-) halos at all output redshifts for the CS-8 run.

We show results for the SMFs at redshifts from $z=6$ to $z=14$ in Fig.~\ref{fig:SMF}. The stellar masses are corrected for numerical resolution as described in Sect.~\ref{subsec:Resolution_correction}. Additionally, we plot the Schechter function fits \cite[]{1976Schechter} of the SMF from \citetalias{2023Kannan-MTNG} as a further $\Lambda$CDM reference, cf. Sect.~\ref{subsec:Resolution_correction}. This provides not only a more accurate baseline, but is also used as a reference for comparison to our CS-8 results at $z=14$, where no halos or galaxies were identified in our CS-10 and $\Lambda$CDM runs as mentioned above.

We additionally indicate the median stellar mass of halos in the first 95\% complete halo mass bin (see Fig.~\ref{fig:HMF}), i.e. the typical stellar mass of galaxies within the least massive well-resolved halos. As galaxies with lower stellar masses are expected to be increasingly affected by the finite numerical resolution, this gives a rough indication of the typical minimum stellar mass of well-resolved galaxies. However, we note that this resolution limit is a more indirect approximation than that shown for the HMFs due to the scatter in the stellar-to-halo mass relation (since the completeness criterion itself is based on a minimum \textit{halo} mass, which we associate with an approximately corresponding stellar mass by considering the median stellar mass of such halos).

For comparison with observational \textit{JWST} data, we also plot the SMF estimates from \citet{2024Weibel}, \citet{2024Navarro-Carrera}, and \citet{2025Harvey}, and indicate the stellar mass of the spectroscopically confirmed, extremely massive $z \sim 14$ galaxy JADES-GS-z14-0 \citep{2024Carniani} in the lower right panel.
Where applicable, these values are converted to a \citet{2003Chabrier-IMF} IMF based on the factors given in \citet[]{2014Madau-Dickinson} to be consistent with our simulations.
We stress that all shown observational SMF estimates are based on SED fits of photometric \textit{JWST} NIRCam \citep{2005Rieke-NIRCam} observations and therefore still currently subject to significant systematic uncertainty (see Sect.~\ref{sec:Introduction}).

Notably, the SMF results from our cosmic strings runs are again converging with our $\Lambda$CDM results towards lower redshifts, so that all simulated SMFs are largely consistent at $z=6$. At higher redshifts, the CS-8 runs predict a significantly larger population of high-mass galaxies than the $\Lambda$CDM run and the $\Lambda$CDM reference simulations from \citetalias{2023Kannan-MTNG}. However, similarly to the HMFs, the SMFs from the CS-10 runs are generally consistent with the $\Lambda$CDM results even at higher redshifts, with only a minor difference at the high-mass end at $z=11$ and a statistically insignificant difference with respect to the Poisson errors at $z=12$.

At redshifts $z=6$ and $z=8$, the observational SMF estimates are broadly consistent with our simulation results, though the \citet{2024Navarro-Carrera} values slightly exceed our predictions at $z=6$, which agree more closely with the \citet{2024Weibel} results. While the \citet{2025Harvey} and \citet{2024Navarro-Carrera} data at $z=8$ fit our cosmic strings and $\Lambda$CDM SMFs roughly equally well, the \citet{2024Weibel} estimates are somewhat more consistent with our $G\mu=10^{-8}$ results at medium-to-high stellar masses, agreeing with them within the cited error bars.

At $z=9$, the preference of the \citet{2024Weibel} data points towards the $G\mu=10^{-8}$ results becomes even more clear, though we note that these estimates are only upper limits at the high-mass end, as indicated by the gray arrows in the plot. However, the \citet{2025Harvey} data instead match our $\Lambda$CDM and $G\mu=10^{-10}$ results noticeably better at this redshift, being just barely consistent with the $G\mu=10^{-8}$ SMF within their respective uncertainties.
Comparing the \citet{2025Harvey} estimates to our results and \citetalias{2023Kannan-MTNG} at $z=11$ and $z=12$ does not show a conclusive preference towards either $\Lambda$CDM / $G\mu=10^{-10}$ or $G\mu=10^{-8}$, as key data points at the high-mass end lie in between the values predicted by the simulations. Only data at lower stellar masses show a preference towards $G\mu=10^{-8}$ at $z=11$ and towards $\Lambda$CDM / $G\mu=10^{-10}$ at $z=12$. However, we caution that our results are likely to be systematically affected by the finite numerical resolution in this mass range, as indicated in the plots by the dash-dotted lines.

Lastly, we show the comparison of our simulated SMF at $z=14$ to the $\Lambda$CDM estimate of \citetalias{2023Kannan-MTNG} at the stellar mass of the massive observed galaxy JADES-GS-z14-0, i.e. $\log (M_{\rm{stellar}} / {\rm M_\odot}) = 8.6^{+0.7}_{-0.2}$ \citep{2024Carniani}. We find that the CS-8 runs predict galaxies with this stellar mass to be approximately 3.5 orders of magnitude more abundant at this high redshift. However, we note that the shown \citetalias{2023Kannan-MTNG} curve is based on results at their closest available output redshift, i.e. $z_{\rm K23} \sim 15$; for reference, the difference to our $z=14$ prediction at the same stellar mass from their next-lower available redshift ($z_{\rm K23} \sim 12$) SMF is roughly 1.5 orders of magnitude.

\subsection{UV luminosity functions}
\label{subsec:UVLF}

\begin{figure*}[!t]
    \centering
    \includegraphics[width=17cm]{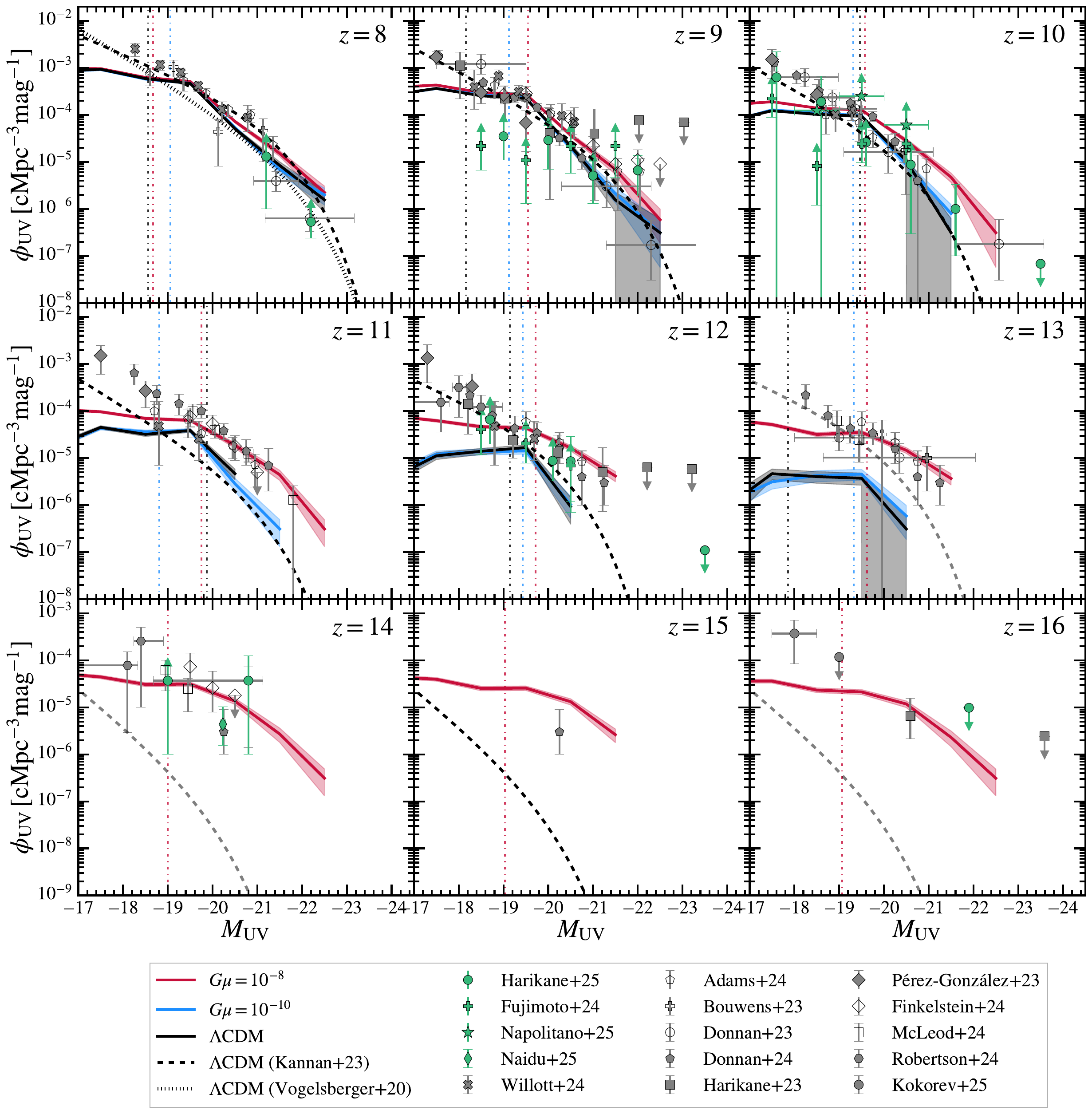}
    \caption{Dust-attenuated UV luminosity functions from our baseline $\Lambda$CDM run (solid black curves) and modified runs modeling cosmic strings with string tension $G\mu=10^{-8}$ (solid red curves) and $G\mu=10^{-10}$ (solid blue curves) with corresponding Poisson errors (shaded error bands) at redshifts $z=8-16$.  
    Dash-dotted vertical lines in the corresponding colors indicate the median UV magnitude of galaxies in the first 95\% complete halo mass bin (see Sect.~\ref{subsec:Results_HMF_SMF} for details).
    At significantly fainter magnitudes than this, the simulation results are expected to be strongly affected by resolution artifacts.
    Dashed and dotted black curves show Schechter fits of the dust-attenuated UV luminosity functions from higher-resolution and larger-volume $\Lambda$CDM simulations using the same code base and galaxy formation model as our runs (\protect\citetalias{2023Kannan-MTNG}; \protect\citealt{2020Vogelsberger-IllTNG_JWST}). We note that we show data from the closest available \protect\citetalias{2023Kannan-MTNG} output redshift $z_{\rm K23}$ for $z=13$ ($z_{\rm K23} \sim 12$), as well as for $z=14$ and $z=16$ ($z_{\rm K23} \sim 15$) and indicate these redshift discrepancies with a reduced opacity of the dashed curve.
    Green and gray symbols show observational estimates from spectroscopically confirmed \protect\citep{2024bHarikane, 2025Harikane, 2024Fujimoto, 2025Napolitano, 2025Naidu} and photometric \protect\citep{2023Bouwens, 2023Donnan, 2023aHarikane, 2023Perez-Gonzalez, 2024Adams, 2024Donnan, 2024Finkelstein, 2024McLeod, 2024Robertson, 2024Willott, 2025Kokorev} \textit{JWST} data, respectively. Where applicable, these estimates are plotted at the closest integer redshift. At high redshifts $z \gtrsim 11$, the observations appear to be in significantly better agreement with the $G\mu=10^{-8}$ results than the $\Lambda$CDM and $G\mu=10^{-10}$ predictions.}
    \label{fig:UVLF}
\end{figure*}

For further comparison with observational data, we show results for our simulated UVLFs in Fig.~\ref{fig:UVLF}. These are obtained by summing up the radiation output at the rest-frame wavelength $1500 \, \r{A}$ in the simulation volume based on BPASS version 2.2.1 tables \citep{2017Eldridge}. We correct for numerical artifacts caused by the sparse sampling of the SFH in low-mass halos close to the simulation's resolution limit by employing the procedure described in \citet{2022Kannan-THESAN}. Specifically, we smooth the age and mass of stars formed fewer than $5\, {\rm Myr}$ ago over the timescale
\begin{equation}
    t_{\rm smooth} = \sum \frac{M_{\rm star} (< 5\, {\rm Myr})}{\rm SFR_{gal}} \ ,
    \label{eq:t_smooth}
\end{equation}
if the corresponding halo satisfies $t_{\rm smooth} > 5\, {\rm Myr}$. Here, ${\rm SFR_{gal}}$ is the instantaneous star formation rate (SFR) of the host galaxy, computed using the sum of the star formation probabilities of all cells on the Equation of State \citep[EoS;][]{2003SpringelHernquist}.
This approach is motivated by the probabilistic star formation routine used in the physics model -- which spawns new star particles stochastically and therefore only occasionally in the lowest-mass halos -- and allows a more robust prediction of the UVLF \citep{2022Kannan-THESAN, 2023Kannan-MTNG}.

To account for the impact of dust grains, we adopt the empirical dust-attenuation ($A_{\rm UV}$) model described in \citetalias{2023Kannan-MTNG}, which is based on the IRX-UV relation given in \citet{2016Bouwens} from Atacama Large Millimeter Array \citep[ALMA;][]{2009Wootten-ALMA} observations at $z \sim 4-7$. Finally, we apply the numerical resolution correction described in Sect.~\ref{subsec:Resolution_correction} to the resulting UV magnitudes $M_{\rm UV}$.

Figure~\ref{fig:UVLF} shows the dust-attenuated UVLFs from our $\Lambda$CDM, CS-8, and CS-10 runs at redshifts $z=8$ to $z=16$.
We also include the corresponding Poisson errors, as well as Schechter fits of the dust-attenuated UVLF obtained from the combination of higher-resolution and larger-volume $\Lambda$CDM simulations presented in \citetalias{2023Kannan-MTNG}. As another reference from higher-resolution runs also using the same simulation code base and galaxy formation model as ours, we additionally plot \citet{2020Vogelsberger-IllTNG_JWST} results from post-processing dust radiative transfer calculations using their dust model C at $z=8$. 
Furthermore, we indicate the median UV magnitude of the least massive 95\% complete halo mass bin (Sect.~\ref{subsec:Results_HMF_SMF}) as a rough estimate of the faintest, but still numerically well-resolved galaxies; however, we stress that this is merely an indirect approximation, since there is typically considerable scatter in the UV magnitudes of galaxies within one halo mass bin.
Additionally, as no halos are identified by the structure finder in the $\Lambda$CDM and CS-10 runs at $z \geq 14$, we compare the CS-8 UVLFs with the \citetalias{2023Kannan-MTNG} results at these high redshifts.

Particularly with regard to redshifts $z \gtrsim 12$, we also note that the start of star formation tends to be delayed in lower-resolution simulations. Additionally, the latter typically exhibit a suppressed star formation rate density $\rho_{\rm SFR}$ at high redshifts ($z \gtrsim 8$) compared to higher-resolution simulations \citep[see e.g.][Fig.~12]{2022Kannan-THESAN}. We attempt to mitigate this effect by our redshift-dependent resolution correction procedure (Sect.~\ref{subsec:Resolution_correction}; see also \citet{2018Pillepich-IllTNG}; \citetalias{2023Kannan-MTNG}), but stress that limited resolution, particularly a delayed onset of star formation, cannot typically be fully corrected for by this prescription (see App.~\ref{app:resolution_tests} for a more detailed investigation of resolution in our simulations).
This is potentially part of the reason for the mismatch between our simulated $\Lambda$CDM UVLF and the $\Lambda$CDM reference from \citetalias{2023Kannan-MTNG} at $z=13$, since it is expected to systematically lower the overall abundance of galaxies in the simulation volume at very high redshifts close to the onset of star formation.\footnote{We note that while this may affect the $\Lambda$CDM and $G\mu=10^{-10}$ results at $z=13$, as this is the first redshift for which we find resolved galaxies in these runs, this is not the case for the $G\mu=10^{-8}$ simulations due to the much earlier ($z > 19$) formation of (resolved) galaxies in the latter.}
Further, we stress that this redshift, along with $z=14$ and $z=16$, is not among the \citetalias{2023Kannan-MTNG} output redshifts $z_{\rm K23}$ for which their UVLF results are available. Therefore, the corresponding plots instead show data from the closest $z_{\rm K23}$, i.e. $z_{\rm K23} \sim 12$ for the $z=13$ panel and $z_{\rm K23} \sim 15$ for the $z=14$ and $z=16$ panels.

We compare our results to observational UVLF estimates based on both spectroscopically confirmed \citep{2025Harikane, 2024Fujimoto, 2025Napolitano, 2025Naidu} and photometrically selected \citep{2023Bouwens, 2023Donnan, 2023aHarikane, 2023Perez-Gonzalez, 2024Adams, 2024Donnan, 2024Finkelstein, 2024McLeod, 2024Robertson, 2024Willott, 2025Kokorev} $\textit{JWST}$ galaxies.

Similarly to the HMFs and SMFs, the $G\mu=10^{-8}$ UVLFs have a significantly higher amplitude at high redshifts compared to the UVLFs inferred from the other runs. The results show the most marked differences at the UV-bright end, while converging with the $\Lambda$CDM and $G\mu=10^{-10}$ results towards lower redshifts. Specifically, by redshift $z=8$ all of our runs' UVLFs are in good agreement with each other, as well as with the $\Lambda$CDM estimates from \citetalias{2023Kannan-MTNG} and \citet{2020Vogelsberger-IllTNG_JWST}.

Additionally, there is again no significant difference between our $G\mu=10^{-10}$ and $\Lambda$CDM UVLFs at any of the considered redshifts, except for a mild increase in the predicted abundance of UV-bright galaxies at $z=11$.

At redshifts $z \lesssim 10$, all of our predicted UVLFs are generally consistent with observational estimates within the cited error bars, though the $G\mu = 10^{-8}$ UVLFs begin to show a mild increase at the bright end by $z=10$.
Notably, as this increase becomes more pronounced at the higher redshifts $z=11$ to $z=13$, the \textit{JWST} data show a significant preference towards the $G\mu = 10^{-8}$ results, again most clearly at the bright end of the UVLFs.
At the highest redshifts considered, $z=14$ to $z=16$, the \textit{JWST} estimates -- including those obtained from observations of spectroscopically confirmed galaxies \citep{2024bHarikane, 2025Harikane, 2025Naidu} -- clearly show significantly better agreement with the $G\mu = 10^{-8}$ predictions than the $\Lambda$CDM baseline.
The remarkable agreement with observational data throughout this broad range of redshifts highlights the $G\mu = 10^{-8}$ cosmic string model's potential to produce a galaxy population consistent with both the highest-redshift \textit{JWST} observations to date and with existing constraints at lower redshifts.

\subsection{Stellar-to-halo mass relation and concentration-mass relation}
\label{subsec:SHMR_cMr}

\begin{figure*}
    \centering
    \includegraphics[width=17cm]{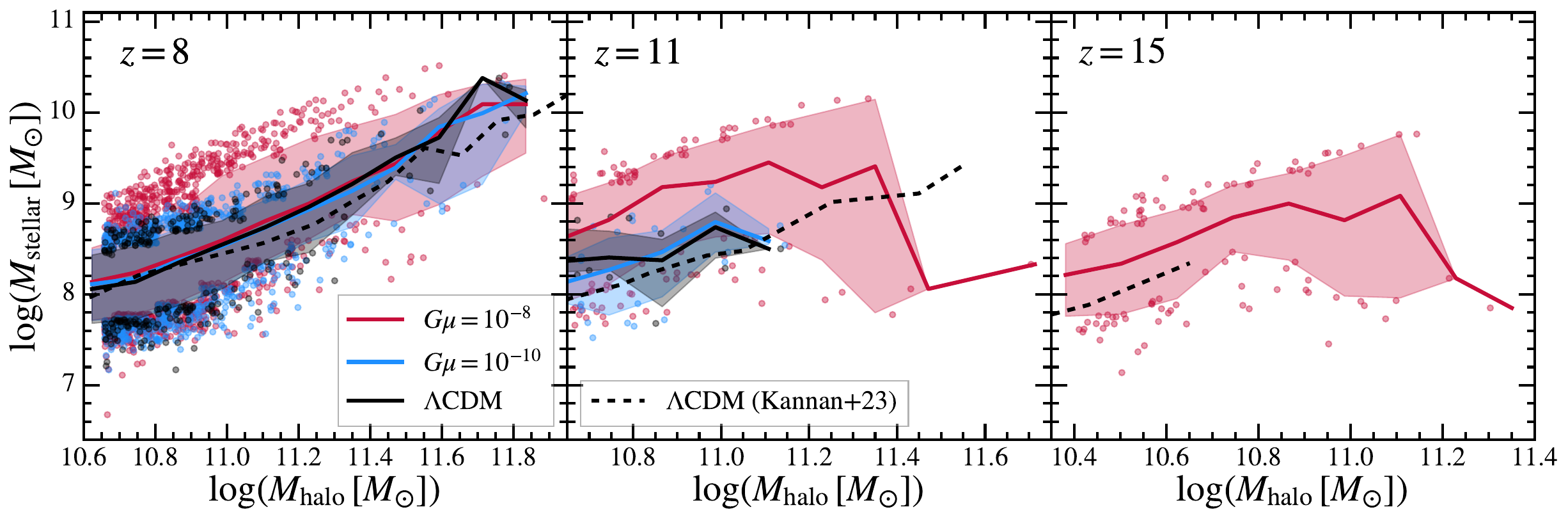}
    \caption{Stellar-to-halo mass ratio of individual simulated galaxies (shaded circles) and median stellar-to-halo mass relation (solid curves) with the 10th to 90th percentile of the distribution (shaded error bands) for our $\Lambda$CDM (black), string tension $G\mu=10^{-8}$ (red), and $G\mu=10^{-10}$ (blue) runs. Shaded circles are only shown for the galaxies outside each mass bin's shaded error bands to improve readability of the figure. We note that the plots include galaxies from all three separate simulations for each of the string tension values and the single baseline $\Lambda$CDM run (cf. Table~\ref{tab:simulations_overview} and Sect.~\ref{sec:Methods}). Therefore, the absolute number of $G\mu=10^{-8}$ and $G\mu=10^{-10}$ scatter points is increased by a factor of three relative to the number of $\Lambda$CDM points. We only include halos with masses in or above the first 80\% complete halo mass bin (see text for details). Dashed black curves show the stellar-to-halo mass relation from \protect\citetalias{2023Kannan-MTNG} as a further $\Lambda$CDM reference.}
    \label{fig:SHMR}
\end{figure*}

We present results for the SHMR and cMr of halos in our simulations in the following, and discuss potential implications in the context of the differences found for the SMFs and UVLFs in previous subsections. Figure~\ref{fig:SHMR} shows the SHMR of galaxies from our $\Lambda$CDM, CS-8, and CS-10 runs. For each model, we additionally plot the median of the SHMR, computed in 20 logarithmically spaced halo mass bins in the mass range $10.2 < \log (M_{\rm halo} / {\rm M_\odot}) < 12.5$, as well as the 10th to 90th percentile of the distribution. Further, we show the simulated SHMR from \citetalias{2023Kannan-MTNG} as an additional $\Lambda$CDM baseline.

In order to remove numerically poorly resolved low-mass halos from our simulations in Fig.~\ref{fig:SHMR}, we only consider halos with a mass equal to or above the first 80\% complete halo mass bin, i.e. the lowest mass bin in which at least 80\% of simulated halos have at least one stellar particle, and the corresponding galaxies. Within the remaining bins, we then only include the individual halos with at least one stellar particle. For this procedure, we use 20 logarithmically spaced bins over the entire halo mass range, $7.5 < \log (M_{\rm halo} / {\rm M_\odot}) < 12.5$. This is a significantly more stringent resolution criterion than only discarding halos with no stellar particles, used to ensure proper convergence of the median SHMR (see also \citealt{2023Yeh}; \citetalias{2023Kannan-MTNG}). However, we note that our 80\% completeness cutoff is set to a slightly lower threshold than the 95\% completeness indicated in previous sections, or the 100\% criterion used in \citetalias{2023Kannan-MTNG}, as a reasonable compromise between showing our results in a broader mass range and the stringency of the resolution criterion.

Similar to previous results, we find no significant difference between the $\Lambda$CDM and $G\mu=10^{-10}$ median SHMR or its scatter, while the $G\mu=10^{-8}$ results do show notable differences. Additionally, by redshift $z=8$, the median SHMR of all runs have converged and are in good agreement with the \citetalias{2023Kannan-MTNG} predictions, though we find a mildly increased scatter in the distribution for the CS-8 runs. 
In particular, there appears to be an increased population of individual massive galaxies within halos of moderate mass in these runs. We find that almost all $G\mu=10^{-8}$ galaxies with masses above the 90th percentile (i.e., above the shaded red error band in Fig.~\ref{fig:SHMR}) correspond to halos seeded by cosmic string loops. However, the relative abundance of these halos is too low to substantially affect the $G\mu=10^{-8}$ median SHMR (red curve).

At higher redshifts $z \gtrsim 11$, we find a significantly increased median SHMR in the $G\mu=10^{-8}$ results across the mass range, with the notable exception of the most massive halos.
Though this extremely massive end of the halo mass distribution ($M_{\rm halo} \gtrsim 11.4$ at $z=11$) is only sparsely sampled within our box volume, we note the presence of individual galaxies with very low stellar-to-halo mass fractions in this range. Tracing these galaxies back in time reveals an onset of star formation at typically very high redshifts $z \gtrsim 19$, notably followed by a depletion of the galaxies' gas reservoirs at exceptionally early times and a complete lack of star formation activity thereafter. We plan to further investigate the evolution of these galaxies and their unique star formation histories in more detail in future work.

Similar individual galaxies are also present at $z=15$ in our $G\mu=10^{-8}$ results. However, in the mass region of overlap with the \citetalias{2023Kannan-MTNG} data, the $G\mu=10^{-8}$ median SHMR is again slightly higher than these $\Lambda$CDM-based predictions.

Our results for the SHMR suggest an overall increased efficiency of star formation in halos found in the CS-8 runs at high redshifts compared to $\Lambda$CDM or CS-10 halos of a similar mass.
To investigate a potential reason for this, we turn our attention to the cMr of our simulated halos.
We fit the density distribution $\rho$ of our simulated dark matter halos as a function of radius $r$ to the Navarro–Frenk–White \citep[NFW;][]{1996NFW} density profile
\begin{equation}
    \rho_{\rm NFW}(r) = \frac{\rho_c}{\frac{r}{r_s} \, \left( 1+\frac{r}{r_s} \right) ^2}
    \label{eq:NFW_profile}
\end{equation}
in log-log-space, with the characteristic density $\rho_c$ and scale radius $r_s$ as fit parameters. From this, we compute the concentration
\begin{equation}
    c \equiv \frac{r_{200}}{r_s} \thickspace,
    \label{eq:NFW_concentration}
\end{equation}
where $r_{200}$ is the radius within which the halo's mean density is equal to 200 times the critical background density $\rho_{\rm crit}$. The concentration parameter quantifies the compactness of halos, with a larger concentration indicating a steeper density profile and, in turn, a more compact halo.

\begin{figure*}
    \centering
    \includegraphics[width=17cm]{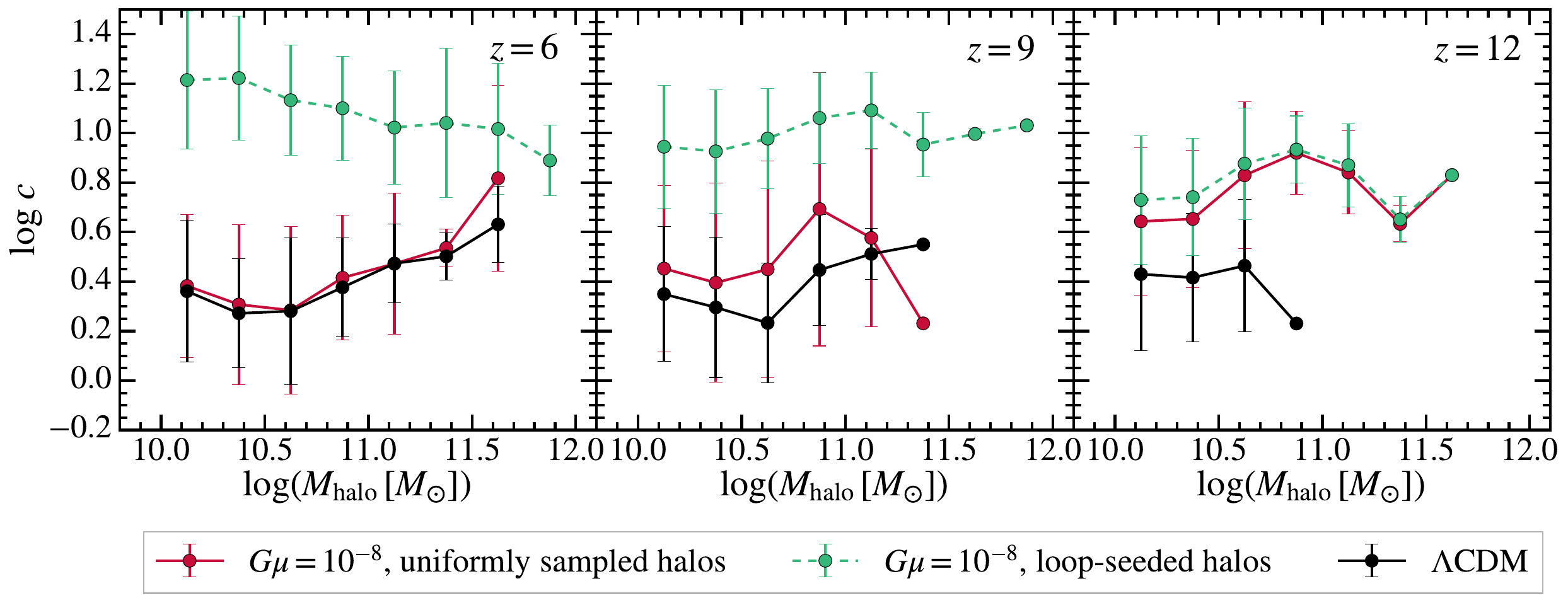}
    \caption{Concentration-mass relation (cMr) of simulated dark matter halos, inferred from a spatially uniform sample of 500 halos in the $\Lambda$CDM (black curves) and CS-8-0 ($G\mu=10^{-8}$, red curves) runs, as well as from CS-8-0 halos seeded by cosmic string loops (dashed green curves), at redshifts $z=6$, 9, and 12. To ensure that individual halos are numerically sufficiently resolved, we only include halos with $M_{\rm halo} \geq 10^{10} \, {\rm M_\odot}$. Filled circles and error bars indicate the mean and standard deviation of the logarithm of the concentrations $\log c$, respectively, in the corresponding mass bins.}
    \label{fig:cMr}
\end{figure*}

In order to estimate the cMr, we sample 500 of the simulated halos with masses well above the resolution limit, $M_{\rm halo} \geq 10^{10} \, {\rm M_\odot}$, uniformly throughout the simulation volume for the $\Lambda$CDM and CS-8-0 runs\footnote{We note that we only sample one of the three CS-8 runs, as it is expected to be representative of all runs with the same string tension; see Appendix~\ref{app:Distribution_CS}.}, respectively, at each output redshift. Further, we sample all halos with $M_{\rm halo} \geq 10^{10} \, {\rm M_\odot}$ seeded by the gravitational effects of cosmic string loops in the CS-8-0 run. We identify these loop-seeded halos by their proximity to the coordinates of cosmic string loops in the simulation volume at the first output redshift $z=19$ and trace their corresponding FOF group forward in time through our snapshots.
We divide the sampled halos into 8 logarithmically spaced mass bins in the range $10 < \log(M_{\rm halo}/{\rm M_\odot}) < 12 $ and compute the mean and standard deviation of the logarithm of the concentration parameter $\log c$ in each bin. The resulting cMr at redshifts $z=6$, 9, and 12 is shown in Fig.~\ref{fig:cMr}.

Across the entire redshift range, we find that the mean concentrations of loop-seeded halos are significantly larger than those of $\Lambda$CDM halos in all mass bins, indicating that loop-seeded halos tend to be more compact. Additionally, we observe that the concentration of loop-seeded halos tends to increase over time.

Further, we find that at $z \gtrsim 12$, the cMr of loop-seeded and uniformly sampled CS-8-0 halos are remarkably similar, which indicates that loop-seeded halos dominate the halo population in the CS-8-0 run at high redshifts. On the other hand, at lower redshifts, the mean concentration of all CS-8-0 halos is considerably lower than that of only the loop-seeded halos; by $z=6$, the $\Lambda$CDM halos clearly dominate the uniform sample. We note that this is consistent with the redshift evolution of both the simulated (Sect.~\ref{subsec:Results_HMF_SMF}) and the analytically predicted \citep{2023JiaoHao-JWST} impact of cosmic string loops on the HMFs.

There are two main factors likely contributing to the larger concentration of loop-seeded halos. Firstly, as they are seeded by point-like sources, loop-seeded halos are expected to have steeper density profiles than those evolved from spatially extended $\Lambda$CDM fluctuations \citep[e.g.][]{mo2010galaxy}, which we find to be the case in our simulations.
Additionally, since the concentration parameter of a halo tends to increase over time, roughly in proportion to $a/a_i$ for a halo initially formed at the time corresponding to the scale factor $a = a_i$ \citep{2002Wechsler-cMr}, the larger concentrations of loop-seeded halos are indicative of earlier formation times.\footnote{We note that this approximate scaling relation is obtained for halos forming and evolving in a $\Lambda$CDM cosmology, and does not necessarily generalize to other cosmological models. We therefore explicitly investigated the evolution of loop-seeded halos in our $G\mu = 10^{-8}$ runs, and similarly found the concentration to grow roughly in proportion to the scale factor.} This is consistent with the results of our CS-8 simulations, as they are the only runs for which we find resolved halos beyond redshift $z=14$, in addition to the HMFs showing a significantly larger population of halos at the high redshifts $9 \lesssim z < 14$.

In light of these results, an interesting question is whether the higher concentration of loop-seeded halos contributes to their larger median SHMR, as higher concentrations could lead to more rapid gas cooling in their dense inner regions, boosting the initial star formation rate in loop-seeded halos.
This may be a potential mechanism producing the highly massive and UV-luminous galaxies found in our simulation volume (Sects.~\ref{subsec:Results_HMF_SMF} and \ref{subsec:UVLF}), or -- somewhat more speculatively -- the galaxies found by \textit{JWST}. We plan to investigate the mass assembly histories of loop-seeded halos and the star formation histories of their galaxies in detail in future work.

Additionally, the increased concentration we find for these halos is particularly intriguing given the recent detection of a million-solar-mass, highly compact object at lower redshift ($z\sim 0.881$) using gravitational imaging \citep{2025Powell_grav_im, 2026Vegetti_grav_im_DM}. While the simulations presented in this work focus on the high-redshift universe, and cannot directly model substructure at this mass scale given our numerical resolution, we note the qualitative similarity of our results to the fact that this object's inferred concentration parameter $c_{\rm vir}$ also appears to significantly exceed $\Lambda$CDM predictions \citep{2026Vegetti_grav_im_DM}.

\section{Discussion and conclusion}
\label{sec:Discussion_and_Conclusion}

In this work, we have presented a suite of large-volume hydrodynamical simulations to numerically investigate the impact of cosmic string loops with string tensions $G\mu = 10^{-8}$ and $G\mu = 10^{-10}$ on the predicted halo and galaxy population at high redshifts.
We modify the initial conditions for the simulations, using the Zel'dovich approximation to compute offsets in dark matter particle positions and velocities due to the gravitational effects of cosmic string loops at $z=127$.
Each of the modified initial conditions then serve as the starting point for a separate full-physics simulation using the \textsc{arepo} code and IllustrisTNG galaxy formation model. As a $\Lambda$CDM baseline, we additionally employ a run based on our unmodified initial conditions, as well as results from previous simulation efforts (\citetalias{2023Kannan-MTNG}; \citealt{2020Vogelsberger-IllTNG_JWST}) sharing the same physics model.

We present and compare our results from the simulations with different cosmic string tensions with these $\Lambda$CDM predictions; specifically, we show high-redshift ($z \geq 6$) predictions for the simulated halo and stellar mass functions, UV luminosity functions, stellar-to-halo mass relation, and concentration-mass relation. Where applicable, we compare our results to observational estimates from recent $\textit{JWST}$ data. Our key findings and conclusions are summarized as follows.
\begin{enumerate}
    \item Modeling cosmic string loops with string tension $G\mu = 10^{-8}$ results in a substantially larger population of UV-bright galaxies at high redshifts in our simulations. Beyond $z \sim 11$, the UV luminosity functions inferred from these runs are in significantly better agreement with observational \textit{JWST} estimates than $\Lambda$CDM simulation predictions from both this work and the existing literature.
    
    \item Observational stellar mass function estimates appear to slightly favor our $G\mu = 10^{-8}$ results as well, though we note that this comparison is less conclusive, particularly since estimates from different \textit{JWST} surveys currently differ significantly. Further, the observational stellar mass function estimates may be subject to uncertainties in the SED modeling and additional assumptions about the high-redshift IMF.
    
    \item Halos in the $G\mu = 10^{-8}$ simulations show a systematically increased median stellar-to-halo mass ratio at high redshifts $z \gtrsim 11$, indicative of more efficient star formation. The latter is frequently invoked as a potentially necessary modification to galaxy formation models in order to achieve consistency with \textit{JWST} observations \citep[e.g.][]{2023Dekel, 2024Finkelstein, 2024Ceverino, 2024Chworowsky}, typically in the framework of standard $\Lambda$CDM cosmology. Our results show the potential of cosmic string models to naturally produce this effect in simulations, without the need to change the physics of the underlying galaxy formation model.
    
    \item We find that halos seeded by $G\mu = 10^{-8}$ cosmic string loops typically have a higher concentration, determined by a fit to the NFW density profile, than $\Lambda$CDM halos of a similar mass. The higher median concentration of loop-seeded halos indicates that they are more compact and therefore likely to have formed at earlier times. We note that this increased compactness may contribute to the higher stellar-to-halo mass ratio we find at high redshifts; we are planning to investigate this possibility in future work.
    
    \item The cosmic string model with string tension $G \mu = 10^{-10}$ has no significant impact on the halo or galaxy population properties considered in this work; rather, the results from these runs are nearly perfectly consistent with $\Lambda$CDM predictions across all redshifts considered. If a string tension of this order or lower were confirmed as a robust upper limit by future observations, our results would therefore tentatively disfavor the cosmic string model as an explanation for \textit{JWST} observations. However, simulations with higher mass resolution are required to provide more conclusive evidence for or against this explanation (see the discussion on our mass resolution in the context of the $G \mu = 10^{-10}$ runs in Sect.~\ref{subsec:Results_HMF_SMF}).

    \item Notably, the results from all of our runs converge with the $\Lambda$CDM baseline by redshift $z=6-8$. Therefore, we find good agreement with observational \textit{JWST} data at the lower redshifts, irrespective of the string tension.
\end{enumerate}

An important limitation of this work is the relatively low resolution of the simulations due to the high computational cost of full-physics hydrodynamical simulations.
We try to mitigate this effect by appropriately rescaling galaxy properties to higher-resolution simulations performed with the same galaxy formation model and code base (Sect.~\ref{subsec:Resolution_correction}). However, numerical resolution affects the timing and efficiency of the earliest star formation in particular, which cannot be fully corrected for with this procedure.

Further, since the string tension $G \mu = 10^{-10}$ halo and galaxy population properties do not differ significantly from the $\Lambda$CDM results, our findings are not equipped to provide evidence either for or against this particular model.

Future studies are expected to help differentiate the cosmic strings model explored in this work from other proposed explanations for the galaxy properties observed by \textit{JWST} (see Sect.~\ref{sec:Introduction}).
Tighter constraints on the high-redshift stellar mass functions and UV luminosity functions will be made possible by wider-area surveys using e.g. \textit{JWST}'s NIRCam or Euclid's wide-field telescope \citep{2022Euclid}, as well as by complementary observations at longer wavelengths with e.g. \textit{JWST}'s Mid-Infrared Instrument \citep[MIRI;][]{2008Wright-MIRI} or ALMA; for more detailed discussions, see e.g. \citet{2024Weibel, 2025Harikane}.
We stress that insights from these could substantially alter the interpretation of our results. In particular, our results would provide evidence \emph{against} the more energetic $G \mu = 10^{-8}$ cosmic string model if future research established lower values for the stellar mass functions and UV luminosity functions at high redshifts.
Crucially, future studies such as precision CMB \citep{2017Hergt} and $21\,\rm{cm}$-line \citep{2010Brandenberger-21cm, 2021Maibach-21cm} observations will also be equipped to provide stronger constraints on viable cosmic string tension values.

%%%%%%%%%%%%%%%%%%%%%%%%
\begin{acknowledgements}
We thank the anonymous referee for constructive feedback and suggestions that helped to improve the paper.
We additionally thank Robert Brandenberger, Laura Sagunski, Oliver Zier, and Xuejian (Jacob) Shen for helpful discussions and comments.
SMK acknowledges funding from a Goethe University International Lab Visits scholarship and MIT Visiting Student scholarship. HJ is supported in part by an NSERC Discovery Grant to R. Brandenberger, by a Milton Leong Fellowship in Science, and by IBS under the project code IBS-R018-D3. RK acknowledges support of the Natural Sciences and Engineering Research Council of Canada (NSERC) through a Discovery Grant and a Discovery Launch Supplement (funding reference numbers RGPIN-2024-06222 and DGECR-2024-00144) and the support of York University's Global Research Excellence Initiative. This research was enabled in part by support provided by the Shared Hierarchical 
Academic Research Computing Network (SHARCNET; \url{www.sharcnet.ca}) and Digital Research Alliance of Canada (\url{alliancecan.ca}).
Computations were performed on the Niagara and Trillium supercomputers at the SciNet HPC Consortium \citep{2010Loken-SciNet, 2019Ponce-Niagara}. SciNet is funded by Innovation, Science and Economic Development Canada; the Digital Research Alliance of Canada; the Ontario Research Fund: Research Excellence; and the University of Toronto. The higher-resolution runs presented in App.~\ref{app:resolution_tests} were run on the FASRC Cannon cluster supported by the FAS Division of Science Research Computing Group at Harvard University.
\end{acknowledgements}

%%%%%%%%%%%%%%%%%%%%%%%%
\bibliographystyle{aa} % style aa.bst
\bibliography{CS.bib}
%%%%%%%%%%%%%%%%%%%%%%%%

\begin{appendix}
\section{Distribution of cosmic string loops}
\label{app:Distribution_CS}

\FloatBarrier

\begin{figure}
    \resizebox{\hsize}{!}{\includegraphics{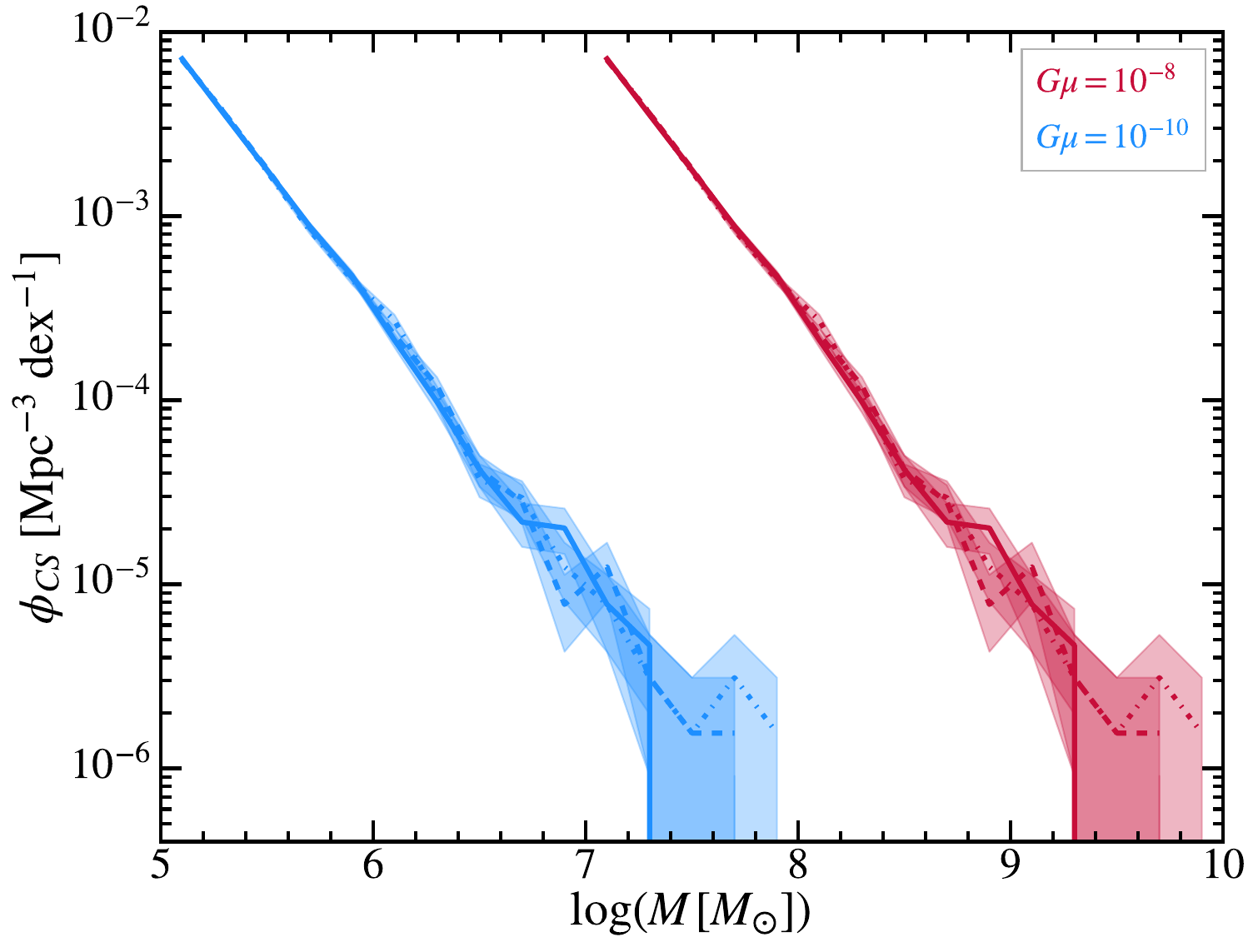}}
    \caption{Mass functions of the different distributions of cosmic string loops used in this work. Solid, dashed, and dash-dotted red curves show the mass functions of loops used in our three string tension $G\mu=10^{-8}$ (CS-8) simulations. Blue curves show the same for our $G\mu=10^{-10}$ (CS-10) runs; the distributions are identical to the $G\mu=10^{-8}$ loops, with masses scaled down by a factor of $1/100$ (see Sect.~\ref{subsec:Implementation_cosmic_strings}). Poisson errors are shown as shaded error bands in the corresponding colors.}
    \label{fig:CS_mass_fcts}
\end{figure}

The number density $n(R, t) \, {\rm d}R$ of cosmic string loops forming at time $t$ in the radius interval between $R$ and $R + {\rm d}R$ in comoving coordinates is given by \cite[]{1994Brandenberger-review}
\begin{equation}
    n(R,t) = 
    \left\{\begin{array}{ll}
        N \alpha^2 \beta^{-2} t_0^{-2} R^{-2} \, , 
            & \alpha t_{\rm eq} \leq R\leq \alpha t\\
        N \alpha^{5/2} \beta^{-5/2} t_{\rm eq}^{1/2} t_0^{-2} R^{-5/2} \, ,
            & R_{\rm cutoff} \leq R\leq \alpha t_{\rm eq}
    \end{array}\right. .
    \label{eq:n_CS_loops}
\end{equation}
Here, $t_{\rm eq}$ is the time of equal matter and radiation, $t_0$ is the present time, and $R_{\rm cutoff}$ is the loop radius corresponding to our cutoff mass (see Sect.~\ref{subsec:Implementation_cosmic_strings}). Further, $N$ is the mean number of long strings crossing any given Hubble volume, determined by the dynamics of cosmic string networks. The parameters $\alpha$ and $\beta$ correspond to the ratios between the loop radius and its formation time, and between the loop length and loop radius, respectively. We assume that the distribution of loops follows the one-scale model, i.e. loops forming at a certain time have the same radius $R = \alpha t$ ($\alpha = \rm const.$). Typical values of $\alpha$ and $\beta$ are determined by Nambu-Goto simulations \citep{1993CS-Scaling, CSsimuls1, CSsimuls2, CSsimuls3, CSsimuls4, CSsimuls5, CSsimuls6, CSsimuls7, CSsimuls8, CSsimuls9, CSsimuls10}. We note that the cutoff radii $R_{\rm cutoff}$ in our simulations are greater than the minimal loop radius due to gravitational radiation, $R_{\rm c}=\Gamma G\mu t$, where $\Gamma$ is the gravitational radiation coefficient \citep{CS-NANO-1, CS-NANO-Wang}.
We sample the radii of cosmic string loops from this number density and compute their masses accordingly based on the value of the string tension $G\mu$ (Sect.~\ref{subsec:Implementation_cosmic_strings}).
The mass functions of the resulting distributions of cosmic string loops are shown in Fig.~\ref{fig:CS_mass_fcts}. 

\begin{figure}
    \resizebox{\hsize}{!}{\includegraphics{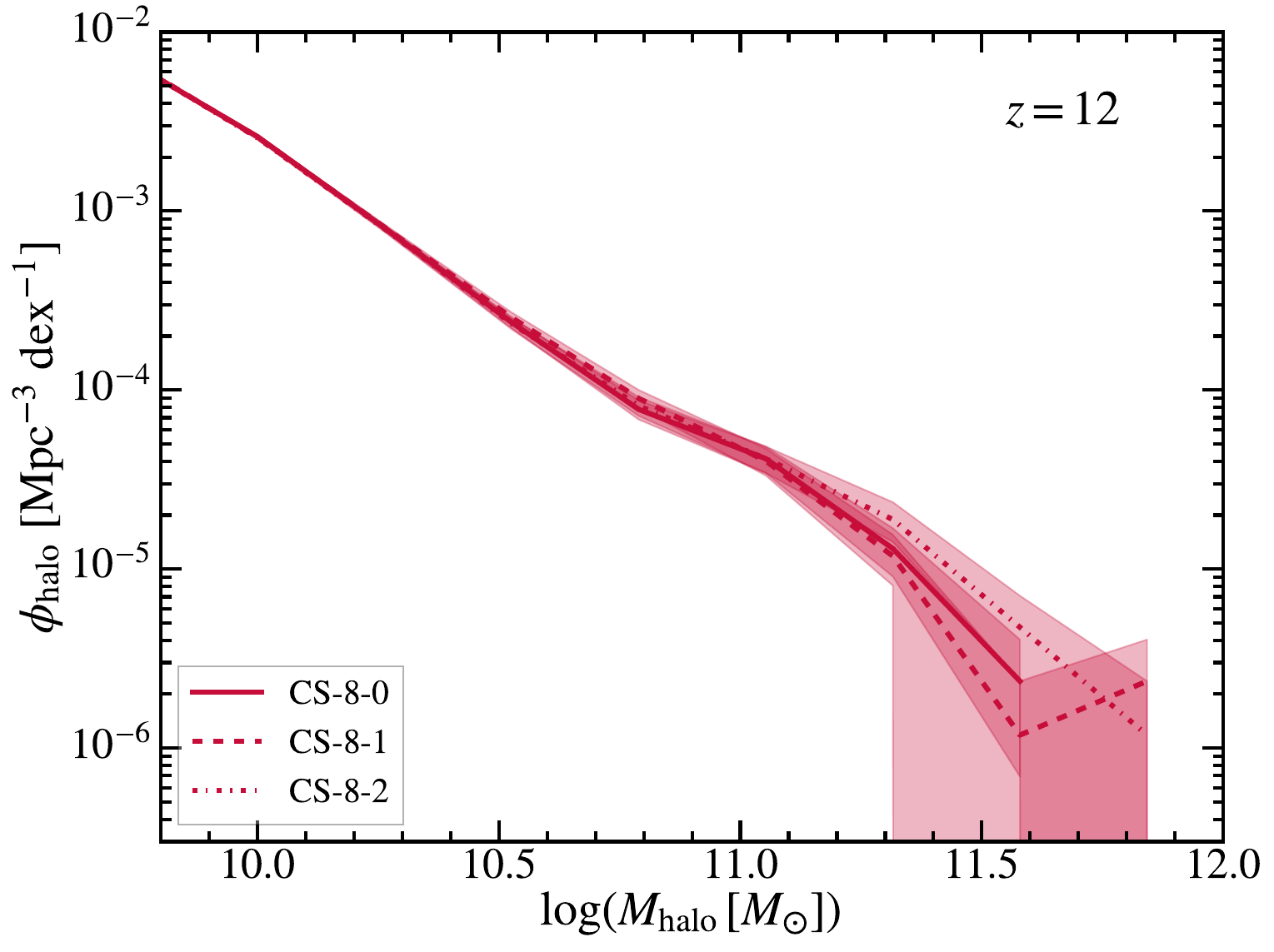}}
    \caption{Halo mass functions computed separately from the CS-8-0 (solid red curves), CS-8-1 (dashed red curves), and CS-8-2 (dash-dotted red curves) runs (see Table~\ref{tab:simulations_overview}). Shaded error bands indicate Poisson errors.}
    \label{fig:HMF_comparison}
\end{figure}

Additionally, Fig.~\ref{fig:HMF_comparison} shows the halo mass functions computed separately from the different $G\mu = 10^{-8}$ runs CS-8-0, CS-8-1, and CS-8-2 (see Table~\ref{tab:simulations_overview}) at $z=12$ as a representative example redshift.
Despite minor differences at the very sparsely sampled highest-mass end, the predicted masses of the halo population are all mutually consistent within their respective Poisson uncertainties. We note that we only include the $G\mu = 10^{-8}$ results in this figure due to the negligible impact of $G\mu = 10^{-10}$ cosmic string loops on the halo mass function (see Fig.~\ref{fig:HMF}). To improve the statistics of our results, we combine outputs from all three runs with string tension $G\mu = 10^{-8}$ and $G\mu = 10^{-10}$, respectively, as described in Sect.~\ref{sec:Results}.

\section{Convergence tests at higher numerical resolution}
\label{app:resolution_tests}

\begin{figure*}
    \centering
    \includegraphics[width=17cm]{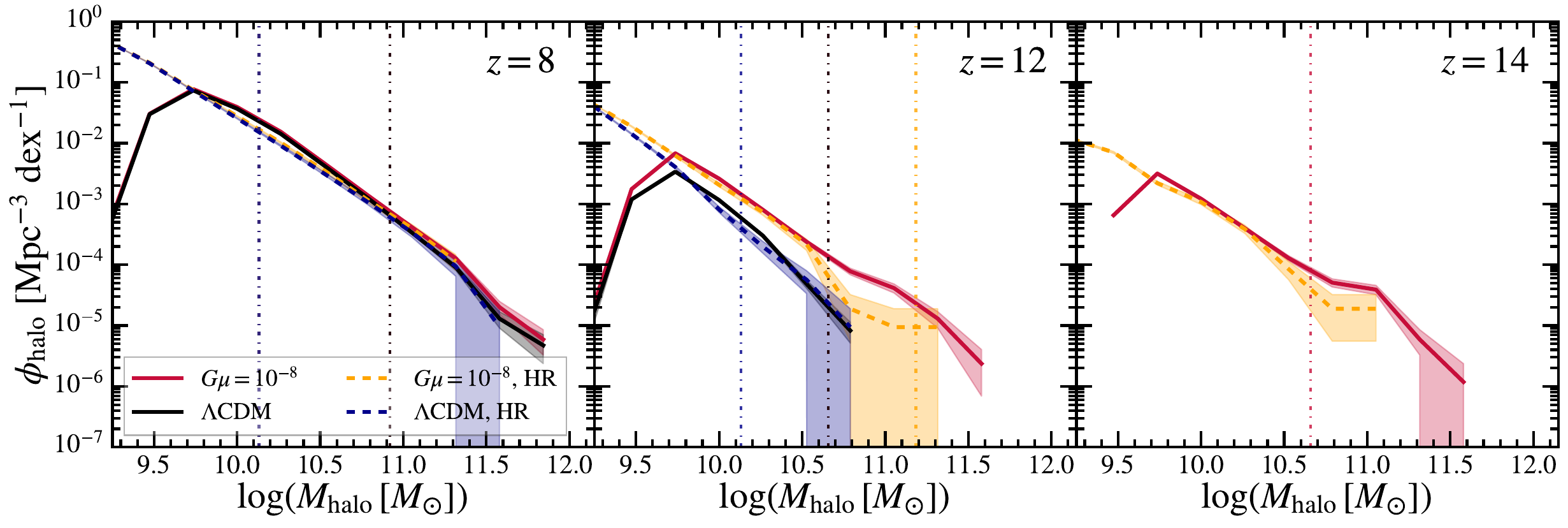}
    \caption{Comparison of halo mass functions inferred from the original runs presented in this paper to those obtained from the higher-resolution (HR) runs performed as a convergence test (see text for details) at redshifts $z=8$, 12, and 14. Solid red and black curves indicate the halo mass functions from the original $G\mu = 10^{-8}$ and $\Lambda$CDM runs, respectively. Dashed orange and dark blue curves show halo mass functions from the HR runs using the $G\mu = 10^{-8}$ and $\Lambda$CDM models, respectively. Shaded error bands in the corresponding colors indicate Poisson errors, while vertical-dash dotted lines indicate the first 95\% complete halo mass bin (see Sect.~\ref{subsec:Results_HMF_SMF}; Fig.~\ref{fig:HMF}).}
    \label{fig:HMF_restest}
\end{figure*}

\begin{figure*}
    \centering
    \includegraphics[width=17cm]{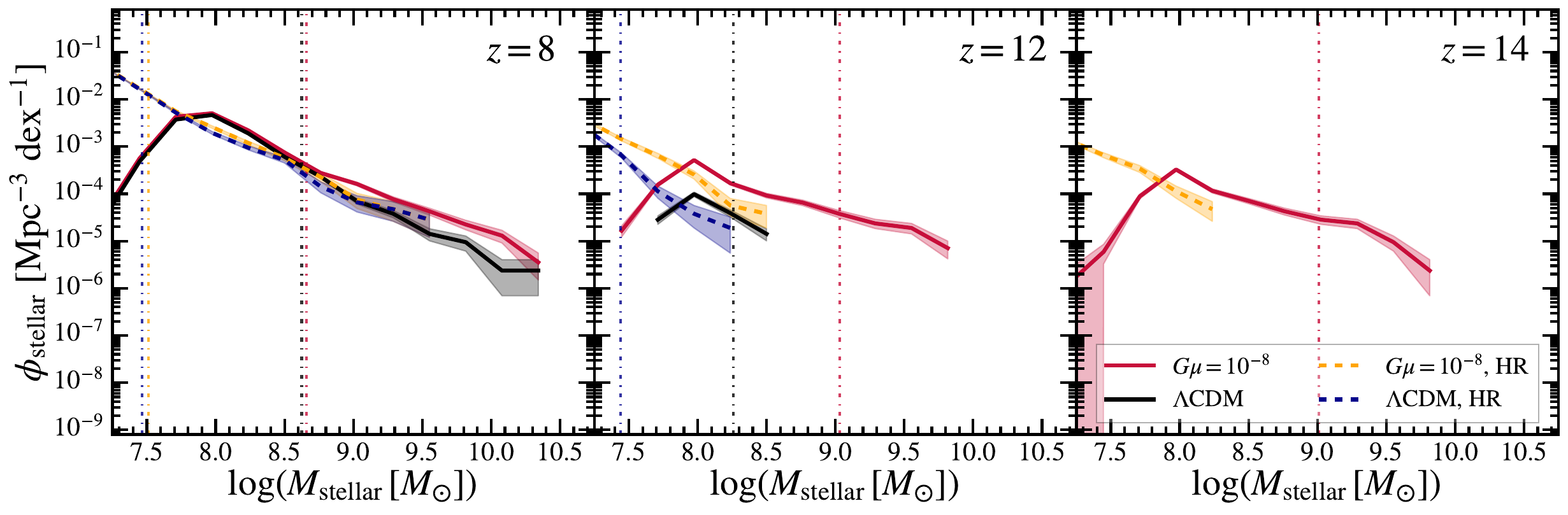}
    \caption{Comparison of stellar mass functions from our original $\Lambda$CDM (solid black curves) and $G\mu=10^{-8}$ (solid red curves) runs to those inferred from higher-resolution (HR) test runs with Poisson errors (shaded error bands) at redshifts $z=8$, 12, and 14. Vertical dash-dotted lines in the corresponding colors indicate the median stellar mass in the first 95\% complete halo mass bin (see Sect.~\ref{subsec:Results_HMF_SMF} for details; cf.~Fig.~\ref{fig:SMF}).}
    \label{fig:SMF_restest}
\end{figure*}

To quantify potential numerical resolution effects of our runs remaining after the resolution correction described in Sect.~\ref{subsec:Resolution_correction}, we compare results of our original simulations described above with additional, higher-resolution runs using the unmodified ($\Lambda$CDM) and string tension $G\mu=10^{-8}$ setups. Due to memory constraints on the machine used, we carried out these runs with the same number of particles as the fiducial runs ($2 \times 850^3$) at half the box size, $L_{\rm box} = 50 \, {\rm cMpc} \ h^{-1} = 74 \,{\rm cMpc}$. This results in an approximately eight times higher mass resolution, with dark matter and gas mass resolutions of $m_{\rm DM} = 2.17 \times 10^7 {\,\rm M_\odot}$ and $m_{\rm gas} = 4.05 \times 10^6 {\,\rm M_\odot}$, respectively. Due to the higher effective resolution, we also modified the gravitational softening length for dark matter and star particles to $2.2 \, \rm ckpc$, as well as the minimum softening length for gas to $0.22 \, \rm ckpc$ (i.e., half the original values; cf.~Sect.~\ref{sec:Methods}). 
Finally, the distribution of the masses and positions of cosmic string loops in the higher-resolution runs with $G\mu = 10^{-8}$ are identical to a $(74 \,{\rm cMpc})^3$ subset of those of the fiducial CS-8-0 run (Sect.~\ref{subsec:Implementation_cosmic_strings}; App.~\ref{app:Distribution_CS}).

We run the higher-resolution (HR) simulations down to redshift $z=8$, and compare their HMFs (Fig.~\ref{fig:HMF_restest}) and SMFs (Fig.~\ref{fig:SMF_restest}) with those from our original runs at $z=8,12$, and $14$ in the following. We apply the same resolution correction procedure as for the main runs (see Sect.~\ref{subsec:Resolution_correction}) to the stellar masses, with the corrections recomputed based on these new runs' resolution, and use the same offsets for both the $\Lambda$CDM, HR and $G\mu = 10^{-8}$, HR runs. Overall, these resolution tests indicate that, in the regime where the resolution and number statistics are sufficiently robust -- i.e., excluding the high-mass end where halo numbers are too few in the smaller-box higher-resolution runs, and excluding the low-mass end strongly affected by the lower resolution of the original runs -- the higher-resolution simulations are mostly consistent with the runs in the main text.

However, at very high redshifts ($z=12$ and 14), we find that this mass range of overlap in which both simulations can produce robust predictions is unfortunately quite small. Due to its smaller simulation volume, the higher-resolution test run can only sparsely sample the rare objects at the high-mass end of the high-redshift halo and galaxy population. Further, the original simulations' lower resolution causes a characteristic dropoff in the mass functions at the poorly resolved low-mass end, as noted in the main text (see e.g.~Fig.~\ref{fig:SMF}). We still present comparisons of our runs at these redshifts to fully show the results (and limitations) of our resolution tests -- performed within the constraints of the computational resources available for this work -- but stress that test runs using a large volume \textit{and} high resolution would be able to put stronger constraints on our runs' convergence at very high redshifts.

Fortunately, at a slightly lower redshift ($z=8$), a much greater mass range can reliably serve as a direct comparison between simulations. At this redshift, we observe good agreement between the runs for both the HMF and SMF.

Further, in the mass range where both simulations are reliable, the HMFs agree well between the higher- and lower-resolution runs at all redshifts considered. We note, though, that the high-mass end illustrates the poor sampling of the rarest objects in the smaller-volume HR runs at $z=12$ and 14, as mentioned above.
At $z=12$, we find that the SMFs appear to have a slight offset between runs, with the higher-resolution runs predicting an overall slightly lower abundance of galaxies in the intermediate mass range overlapping with predictions from the original runs. However, the prediction of a larger abundance of more massive ($M_{\rm stellar} \gtrsim 10^{8} \, M_{\odot}$) galaxies in the $G\mu = 10^{-8}$ HR run compared to the $\Lambda$CDM HR run remains robust.

The small overlap of the sufficiently resolved and sampled mass range is unfortunately particularly apparent at $z=14$ for the SMFs. We therefore do not attempt to draw overly strong conclusions about the convergence of our simulated stellar masses at such high redshifts, but note that we do find that the overall amplitude and slope appear consistent with the $z=14$ SMF obtained from the larger-volume original runs.

\end{appendix}

\end{document}